\documentclass[aps,prb,reprint,twocolumn,showpacs,floatfix,superscriptaddress]{revtex4-2}

\usepackage{amssymb,amsmath,amstext}                
\usepackage{graphicx}                                               
\usepackage{epstopdf}                                               
\usepackage{color}                                                     
\usepackage{bm}                                                        
\usepackage{appendix}                                              
\usepackage[utf8]{inputenc}
\usepackage{latexsym}
\usepackage{xcolor}
\usepackage{braket}
\usepackage{siunitx}
\usepackage{umoline}
\usepackage{epsfig}
\usepackage{graphics}
\usepackage{psfrag}
\usepackage{mathtools}
\usepackage[english]{babel}
\usepackage[T1]{fontenc}
\usepackage{lmodern}
\usepackage{rotating}
\usepackage{multirow}

\definecolor{lblue} {RGB}{51,71,158}
\usepackage[colorlinks=true,citecolor=blue,linkcolor=blue,urlcolor=lblue]{hyperref}

\usepackage{appendix}

\begin{document}
\setcounter{section}{0}

\title{Many-body localization crossover is sharper in quasiperiodic potential 
}  

\author{Pedro R. Nic\'acio Falc\~ao}
\affiliation{Szkoła Doktorska Nauk \'Scis\l{}ych i Przyrodniczych, Uniwersytet Jagiello\'nski,  \L{}ojasiewicza 11, PL-30-348 Krak\'ow, Poland}
\affiliation{Instytut Fizyki Teoretycznej, Wydzia\l{} Fizyki, Astronomii i Informatyki Stosowanej,
Uniwersytet Jagiello\'nski,  \L{}ojasiewicza 11, PL-30-348 Krak\'ow, Poland}

\author{Adith Sai Aramthottil}
\affiliation{Szkoła Doktorska Nauk \'Scis\l{}ych i Przyrodniczych, Uniwersytet Jagiello\'nski,  \L{}ojasiewicza 11, PL-30-348 Krak\'ow, Poland}
\affiliation{Instytut Fizyki Teoretycznej, Wydzia\l{} Fizyki, Astronomii i Informatyki Stosowanej,
Uniwersytet Jagiello\'nski,  \L{}ojasiewicza 11, PL-30-348 Krak\'ow, Poland}

\author{Piotr Sierant}
\affiliation{ICFO-Institut de Ci\`encies Fot\`oniques, The Barcelona Institute of Science and Technology,
Av. Carl Friedrich Gauss 3, 08860 Castelldefels (Barcelona), Spain}

\author{Jakub Zakrzewski} 
\affiliation{Instytut Fizyki Teoretycznej, Wydzia\l{} Fizyki, Astronomii i Informatyki Stosowanej,
Uniwersytet Jagiello\'nski,  \L{}ojasiewicza 11, PL-30-348 Krak\'ow, Poland}
\affiliation{Mark Kac Complex Systems Research Center, Uniwersytet Jagiello{\'n}ski, Krak{\'o}w, Poland}

\date{\today}

\begin{abstract}
The strong disorder may significantly slow down or even completely hinder the thermalization of quantum many-body systems due to many-body localization (MBL). A sufficiently deep quasiperiodic potential may also inhibit thermalization. 
In this work, we numerically demonstrate direct differences in the behavior of standard ergodicity-breaking indicators at the MBL crossover in random and quasiperiodic systems. Our key finding is the exponential increase in the sharpness of the MBL crossover with system size for quasiperiodic systems, a trend that is only linear in disordered systems. The strong tendency towards a non-analytic behavior in quasiperiodic systems is consistent with the existence of dynamical regimes with sharply defined boundaries or an MBL phase transition. It highlights the importance of quasiperiodic systems for our understanding of many-body dynamics.
\end{abstract}

\maketitle

\section{Introduction}

According to the eigenstate thermalization hypothesis~(\textbf{ETH})~\cite{Deutsch91, Srednicki94, Rigol08, Dalessio16}, isolated quantum many-body systems prepared in an out-of-equilibrium state undergo thermalization reaching an equilibrium state in which local observables are described by appropriate ensembles of statistical mechanics~\cite{Foini19, Pappalardi22, Pappalardi23, pappalardi2023microcanonical}. This ergodic paradigm, followed by multiple quantum many-body systems~\cite{Rigol09, Santos10,  Steinigeweg13, Khatami13Fluctuation, Beugeling14, Schonle21lens}, may be broken in the presence of disorder. The disorder slows down the process of thermalization~\cite{Luitz16, Bera17, Sels23dilute, Evers23Internal}, leading to a dynamical regime in which the transport is suppressed \cite{Znidaric16}, and the entanglement spreads slowly~\cite{DeChiara06, Znidaric08, Bardarson12, serbyn2013universal, Iemini16signatures}. Sufficiently strong disorder leads to a phenomenon of the many-body localization (\textbf{MBL})~\cite{Gornyi05, Basko06, Oganesyan07}: the dynamics slow down so dramatically that finite-size systems fail to thermalize~\cite{Pal10, Huse14, Luitz15, Ros15, Serbyn13b, Imbrie16, Wahl17, Mierzejewski18, Thomson18}. Early studies extended this conclusion to the infinite system size limit, claiming a stable MBL \textit{phase}~\cite{Nandkishore15, Imbrie16a, Alet18,  Abanin19}. This interpretation has been, however, questioned~\cite{Suntajs20e} and subsequent investigations~\cite{Sierant20t, Panda20, Sels20obstruction, Kiefer20, Sierant20p, Abanin21, Kiefer21Unlimited, Ghosh22, Sels21} demonstrated the difficulty of distinguishing between extremely slow thermalization and its complete arrest as reviewed in~\cite{Sierant24rev}. Regardless of the status of the MBL phase, numerical and experimental studies find a robust MBL \textit{regime} in which thermalization does not occur on experimentally relevant time scales~\cite{Sierant22c, Morningstar22}.

Besides fermionic~\cite{Mondaini15, Prelovsek16, Zakrzewski18, Kozarzewski18} and bosonic~\cite{Sierant17,Sierant18, Orell19, Hopjan19} lattice systems, the paradigmatic models of MBL include spin-1/2 chains~\cite{Berkelbach10, Serbyn15, Agarwal15, Bera15, Serbyn16, Enss17infinite, Herviou19, Colmenarez19, Sierant20model, Schiulaz20, TorresHerrera20, Colbois23, Colbois24}  with Hamiltonian of the form 
\begin{equation}
    \hat{H} = \hat{H_0} + \sum_{j=1}^L h_j \sigma^z_j,
    \label{eq:dis}
\end{equation}
where $\hat{H}_0$ is a translationally invariant short-range interacting Hamiltonian, $L$ is the system size, $\sigma^{x,y,z}_i$ denote Pauli matrices and the fields $h_i$ are chosen as independent random variables uniformly from the interval $[-W,W]$. The term $\sum_{i=1}^L h_i \sigma^z_i$ breaks the translational invariance and introduces random disorder (\textbf{RD}) of strength $W$ to the system. Another way to break the translational invariance of $\hat{H_0}$ is to introduce a quasi-periodic (\textbf{QP}) potential 
\begin{equation}
    \hat{H} = \hat{H_0} + \sum_{j=1}^L W\cos(2\pi \kappa j +\varphi) \sigma^z_j, 
    \label{eq:qp}
\end{equation}
where $\kappa$ is an incommensurability factor (e.g. $\kappa = \kappa_{\mathrm{gr}} \equiv (\sqrt{5} -1)/2$) and $\varphi$ is a global phase randomly selected from the interval $[0,2\pi]$ for each realization of the 
QP potential. Increase in the amplitude $W$ of the RD and QP potential leads to inhibition of thermalization and the onset of MBL regime~\cite{Pal10, Iyer13, Naldesi16, Setiawan17, BarLev17, Bera17a, Weidinger18, Doggen19, Weiner19, Mace19, Singh21, Agrawal22, Zhang22, Strkalj22, Xu19butterfly, Tu23spectrum, Vu22fermionic, Prasad23SingleParticle, Thomson23}. 
The QP potential is 
realized naturally in optical lattices by imposing a secondary standing wave with a period incommensurate with the primary lattice~\cite{Schreiber15, Luschen17, Lukin19, Rispoli19, Leonard23}. However, the RD and QP potentials are starkly different. The fields $h_i$ for RD are uncorrelated, while the QP potential~\eqref{eq:qp} is deterministically fixed by the phase $\varphi$ for each realization, resulting in non-vanishing correlations between the on-site potential values at arbitrarily distant sites. Consequently, RD systems are more heterogeneous and host Griffiths type~\cite{Vojta10} rare regions of locally weaker disorder~\cite{Gopalakrishnan16, Gopalakrishnan15, Agarwal17, Pancotti18, DeRoeck17, Luitz17bath, Peacock23, Szoldra24, Scocco24propagation}, while QP systems are more uniform~\cite{Gopalakrishnan20}. 

Do the disparities of the RD and QP potential translate to any essential differences in the properties of MBL in the two types of systems? 
{The gross features of the dynamics in the MBL regime of QP systems~\cite{Schreiber15, Luschen17, Lukin19, Rispoli19, Leonard23} and of RD systems~\cite{Smith16, Choi16, Guo21, Chiaro22} are similar. This includes a slow decay of local correlation functions~\cite{Luitz16, Luschen17, Weiner19, Sierant22c}, a logarithmic in time growth of entanglement entropy both in RD ~\cite{DeChiara06, Znidaric08, Bardarson12} and QP~\cite{Lee17, Lukin19} systems, and a long-lived memory of the initial state~\cite{Serbyn13a, Ros15, Singh21}.}
{Moreover, the standard ergodicity breaking indicators such as the average gap ratio probing energy level statistics}~\cite{Oganesyan07, Atas13} {or rescaled entanglement entropy of eigenstates show qualitatively similar behavior at the ETH-MBL crossover both in RD and QP systems}~\cite{Khemani17, Aramthottil21}.
{Differences between RD and QP systems, suggested by a renormalization group schemes}~\cite{Zhang18, Zhang2019strong}{, become apparent when}  quantities measuring sample-to-sample fluctuations of entanglement entropy~\cite{Khemani17}, level statistics~\cite{Sierant19level}, and properties of 
self-energy term in the Fock space propagator~\cite{Ghosh24FockPropagator} are considered. Such quantities
indicate {weaker} fluctuations of the system's properties for QP potential than for RD. Similarly, finite-size collapses point out weaker system size drifts for QP potential~\cite{Aramthottil21}, simulations of time dynamics reveal weak but noticeable long-time oscillations in correlation functions of local observables for QP potential, unobservable for the RD case~\cite{Sierant22c}, {while computations of} diffusion constant {indicate that it} vanishes faster {with growth of QP potential's amplitude than in RD systems}~\cite{Prelovsek23}.

In this work, we ask whether the differences between the QP and RD potentials imply differences in the behavior of the standard ergodicity-breaking indicators, such as the average gap ratio or probes of entanglement of eigenstates at the ETH-MBL crossover. To this end, we perform a large-scale numerical investigation of the ETH-MBL crossover in QP and RD Floquet systems. 
The latter are contrasted
with results for  RD, Eq.~\eqref{eq:dis}, and QP, Eq.~\eqref{eq:qp}, autonomous spin chains. Analyzing QP models, we find persistent drifts of ergodicity-breaking indicators at the ETH-MBL crossover, akin to the RD case. Despite these similarities, there is an immediate disparity between the QP and RD potentials. The ETH-MBL crossover sharpens up exponentially quickly with increasing system size $L$ for QP potential, while the sharpening up is only linear with $L$ for the RD. This demonstrates a direct difference between the standard bulk ergodicity-breaking indicators at ETH-MBL crossover in the RD and QP systems and constitutes the main finding of this work.

\section{Models and Numerical Methods} 
Similarly to the autonomous spin chains, periodically driven Floquet systems may undergo MBL~\cite{Lazarides15, Ponte15, Ponte15a, Abanin16, Zhang16, Bairey17, Sahay21, Garratt21MBLasSymmetryBreaking, Sonner21ThoulessEnergy} which prevents their heating \cite{Moessner17}, enabling creation of driven time crystals \cite{Sacha15, Sacha17,Khemani16structure, Else16, Choi17,Bordia17, Pizzi20} and stabilization of Floquet insulators \cite{Po16, Nathan19, Roy17, Harper17, Rudner20}. We focus on a Kicked Ising Model (\textbf{KIM})~\cite{Prosen02, Prosen07} defined by the Floquet operator
\begin{equation}
    \hat{U}_{\mathrm{KIM}} = e^{-i \sum_{j=1}^L g \sigma_{j}^x}e^{-i\left( \sum_{j=1}^{\ell} J \sigma_j^{z}\sigma_{j+1}^{z} + \sum_{j=1}^{L} h_j\sigma_j^{z} \right) },
    \label{KIM}
\end{equation}
where $h_j$ is the on-site magnetic field 
and $\ell$ determines the boundary conditions of the system.  For QP systems due to the nature of QP potential, we consider open boundary conditions and set 
$\ell=L-1$. For RD we set periodic boundary conditions by putting 
$\ell=L$ and identifying $j=L+1$ with $j=1$.
We have verified that our conclusions apply also to RD systems with open boundary conditions. 
For RD, system size drifts at the ETH-MBL crossover in KIM are milder than in autonomous spin chains~\cite{Sierant23st}. To consider QP KIM, we set $h_j = \pi[1 + \cos(2\pi\kappa j + \varphi)]$ with $\kappa = \kappa_{\mathrm{gr}}$
and select $\varphi \in (0,2\pi]$ with uniform probability. To tune the system across ETH-MBL crossover, we set $J=g=1/W$, with $W$ playing the role of the effective strength of QP potential. Our results for QP KIM will be contrasted with RD KIM, for which the onsite fields are selected with uniform probability from the interval $h_j \in [0, 2\pi]$ and we still follow the choice $J=g=1/W$, with $W$ playing the role of disorder strength for RD KIM. 

To show the generality of our findings, we consider also the Heisenberg spin-1/2 chain~(\textbf{HSC})~\cite{Berkelbach10, Serbyn15, Agarwal15, Bera15, Serbyn16, Enss17infinite, Herviou19, Colmenarez19, Sierant20model, Schiulaz20, TorresHerrera20, Colbois23, Colbois24} by setting 
\begin{equation}
     \hat{H}_{0} = \frac{J}{4}\sum_{j=1}^{\ell} \vec{\sigma}_j \cdot \vec{\sigma}_{j+1}
\end{equation}
in Eq.~\eqref{eq:dis} to arrive at RD HSC and in Eq.~\eqref{eq:qp} to obtain the QP HSC.

To study the ETH-MBL crossover in Floquet (autonomous) systems, we compute the eigenvectors $|\phi_n\rangle$ and the eigenvalues $e^{i\phi_n}$ ($E_n$) using the POLFED algorithm \citep{Sierant20p}. For the Floquet case, the algorithm relies on geometric sum spectral transformation~\cite{Luitz21, Sierant23st}, allowing us to study system sizes up to $L=20$, substantially 
beyond the scope of 
traditional full exact-diagonalization schemes, limited{, on present-day computers,} to $L\leq 14$ for unitary matrices.  

We study the properties of two ergodicity-breaking indicators commonly employed in the numerical analysis of the ETH-MBL crossover~\cite{Luitz14universal, Sierant20p}: the mean gap ratio and the entanglement entropy of eigenstates. The mean gap ratio is defined~\cite{Oganesyan07} as 
\begin{equation}
    \overline{r} = \langle \min(g_i,g_{i+1})/\max(g_i,g_{i+1}) \rangle
\end{equation}
where $g_i = \phi_{i+1} - \phi_i$ is the gap between two consecutive eigenphases for Floquet systems (while $g_i = E_{i+1} - E_{i}$ for HSC), and  $\langle ... \rangle$ represents the average over eigenstates and disorder realizations.
To characterize eigenvectors, we compute their entanglement entropy by splitting the system in two subsystems 
and evaluating 
\begin{equation}
  \mathcal{S} =  -\sum_{i=1}^{i_M} \lambda_{i}^2\log\lambda_{i}^2  
\end{equation}
where $\lambda_i$ are the Schmidt coefficients of the eigenstate $\vert \phi_n \rangle $~\cite{Karol}. The entanglement entropies are averaged over eigenstates and disorder realizations, resulting in $\langle\mathcal{S} \rangle$. Subsequently, we rescale the result by the average entanglement entropy $\mathcal{S}_{\mathrm{COE}}$ of the eigenstates of Circular Orthogonal Ensembles (COE) matrices \cite{Haakebook}, i.e., $\overline{s} = \langle\mathcal{S} \rangle /\mathcal{S}_{\mathrm{COE}}$~\cite{Alessio14, Vidmar17}.

As an additional measure {of entanglement of eigenstates} we use the mean Schmidt gap, $\overline \Delta$, obtained by averaging Schmidt gaps $\Delta= \lambda_1^2-\lambda_2^2$ of eigenstates. One expects the average Schmidt gap to be {decaying exponentially with system size $L$ for the}
{eigenstates in the ETH regime,} while in the MBL regime, with low-entanglement entropy, the mean Schmidt gap is expected to 
be {independent of system size~\cite{Gray18}.}

In the following analysis, we compute $\mathrm{N_{ev}} = \min(2^L/10,1000)$ eigenvectors with eigenphases closest to $\phi_n = 0.0$ for Floquet systems.  For systems up to $L=16$, we consider at least $10^{4}$ realizations of the QP (or RD) potentials for each value of $W$, while for $L=17-18$ and $20$, we take at least $10^3$ and $200$ realizations of the system, respectively. 

\begin{figure*}[t!]
    \centering
    \includegraphics[width=1 \textwidth]{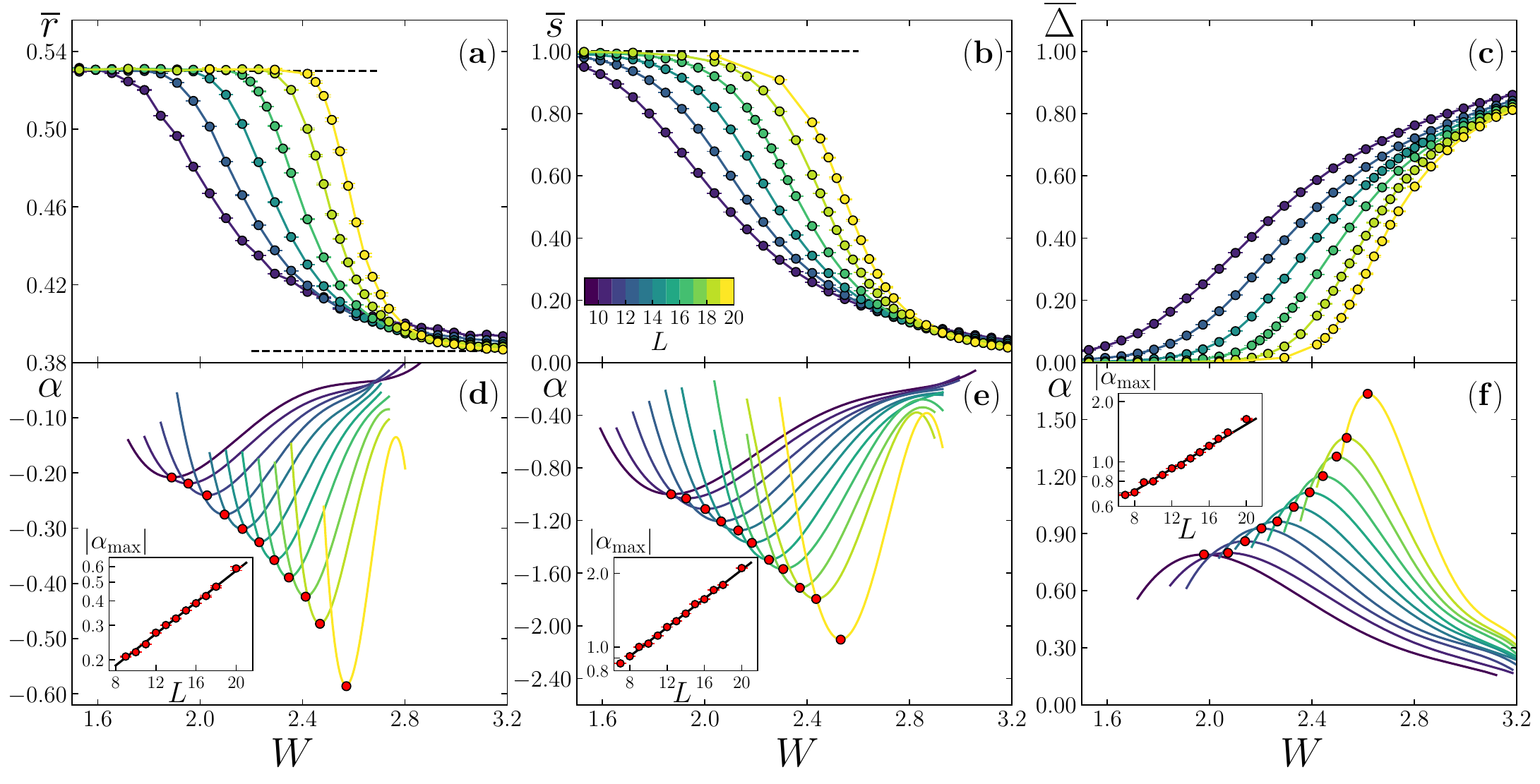}
    \caption{\textbf{Top panel}: The ETH-MBL crossover in QP KIM showed by typical ergodicity breaking indicators as a function of the effective quasi-periodic (QP) potential strength $W$ for various system sizes $L$. These indicators include the (a) average gap ratio $\overline{r}$, (b) the rescaled entanglement entropy $\overline{s}$, and (c) the Schmidt gap $\overline{\Delta}$. \textbf{Lower panel}: Their respective derivatives $\alpha = d{\overline{x}}/dW $ as function of $W$, where $\overline{x}$ is the respective observable.  The red circles highlighting the maximum slope $|\alpha_{\max}|$ for each $L$. Inset: A semi-logarithmic plot of $|\alpha_{\max}|$ versus $L$, illustrating an exponential scaling behavior.}
    \label{fig1}
\end{figure*}

\section{The ETH-MBL crossover in QP Kicked Ising Model}

While Floquet systems with RD were quite extensively studied, as reviewed above, it is not so with QP potential case.
Apart of the early experiment on Floquet MBL in QP potential \cite{Bordia17p} the theoretical studies concentrated mostly on multifractality of eigenstates \cite{Sarkar22,Zhang22}. We shall consider the KIM in QP potential having the data for RD potential at hand \cite{Sierant23st}. With the growth of QP potential strength $W$, the mean gap ratio $\overline r$, shown in Fig.~\ref{fig1}~(a), interpolates between the value $\overline{r}_{\mathrm{COE}} \approx 0.53$ characteristic for ergodic systems~\cite{Alessio14} and $\overline{r}_{\mathrm{PS}} \approx 0.386$, signifying the MBL regime~\cite{Oganesyan07}. In the same disorder range, the rescaled entanglement entropy $\overline{s}$ changes from volume-law ($\overline{s}\approx 1$) to area-law ($\overline{s}\approx 0$)
as shown in Fig.~\ref{fig1}~(b). In the same spirit, the ETH-MBL crossover is also captured by the Schmidt gap $\overline{\Delta}$ which distinguishes the ergodic regime in which $\overline{\Delta}$ decreases with increase of $L$ from the MBL regime at large $W$, where $\overline{\Delta}$ is approximately independent of $L$, as shown in Fig.~\ref{fig1}~(c).  

At first glance, the behavior of these standard ergodicity breaking indicators is similar to the RD case{~\cite{Oganesyan07, Pal10, Luitz15, Gray18, Sierant20p}}. However, the analyzed ETH-MBL crossover in QP KIM also possesses certain unexpected features characteristic for the QP systems. Firstly, $\overline r$ is not changing smoothly with $W$ at small ($L \lesssim 10$) system sizes but rather fluctuates around a smooth curve monotonously decreasing with $W$. The fluctuations can be reduced by introducing randomness to the incommensurability factor $\kappa$, which can be taken from a narrow distribution centered at $\kappa_{\mathrm{gr}}$, see Appendix~\ref{random}. Hence, we fix $\kappa=\kappa_{\mathrm{gr}}$ for $L \geq 12$, while for smaller system sizes we take $\kappa$ uniformly distributed in the interval $[0.98, 1.02] \times \kappa_{\mathrm{gr}}$. The second feature of ETH-MBL crossover characteristic for QP potential is that the crossover significantly  \textit{sharpens up} with increasing $L$.

To quantitatively analyze the sharpening up of the ETH-MBL crossover,  we fit the observable $\overline{x}$ (with $\overline{x} \in\{ \overline r, \overline s, \overline \Delta \} $) in the interval ${\overline{x}} \in [{\overline{x}_{\mathrm{COE}}} - \delta_{{\overline{x}}},{\overline{x}}_{\mathrm{PS}} + \delta_{{\overline{x}}}]$ with a degree $5$ polynomial function of $W$ and evaluate its derivative $\alpha = d{{\overline{x}}}/dW$.
The choice $\delta_{\overline{x}}$ depends on which observable we are looking at. For the average gap ratio, we take $\delta_{\overline{r}} = 0.01$, while for the rescaled entanglement entropy and the Schmidt gap $\delta_{\overline{x}} = 0.1$. We have checked that the results are not too sensitive to the values of $\delta_{\overline{x}}$ chosen, see Appendix~\ref{robustness}.
Interestingly,  the corresponding absolute value of the slope $| \alpha_{\max} |$, we obtain, is increasing, to good accuracy, exponentially with the system size, 
as shown in the insets of the lower panels in Fig.~\ref{fig1}~(d, e, f). Parametrizing the exponential dependence as {$| \alpha_{\max} | = A e^{\lambda L}$,  where $A, \lambda $ are constants, we obtain,}
for the average gap ratio, $\overline{r}$, $\lambda = 0.093(4)$, while for the rescaled entanglement entropy, $\overline{s}$ we get $\lambda = 0.068(2)$ and for the Schmidt gap, $\overline \Delta$, we find $\lambda = 0.064(2)$.

The robustness of the exponential scaling 
{of $|\alpha_{\max}|$} can be tested by looking at different intervals of $W$, polynomial functions, or fixed points of $W$, with all these methods giving a similar exponent $\lambda \approx 0.1$, as we further discuss in the Appendix~\ref{robustness}. {While the uncovered increase of $| \alpha_{\max} | $  is \textit{clearly} faster than linear in $L$, the dependence of $|\alpha_{\max} |$ on $L$ could also be fitted by a second, or higher order polynomial of $L$. However, such fits would involve at least three fitting parameters. In contrast, the exponential system size dependence, $| \alpha_{\max} | = A e^{\lambda L}$, is the simplest, two-parameter formula that accurately describes the behavior of $| \alpha_{\max} | $ for QP KIM in the interval of system sizes accessible to us. With this remark in mind, we keep referring to the behavior of the $| \alpha_{\max} |$ in QP KIM as the exponential growth. 

The behavior of the parameter $\lambda$ is robust against different choices of the disorder strength at which the value of $d\overline{x}/dW$ is analyzed. For instance, instead of considering the {disorder strength corresponding to the} maximum {of the absolute value of the} derivative, let us consider the slopes, {$\alpha$}, at the {disorder strength at which the ergodicity-breaking indicators admit a value} $\overline{x}_{\mathrm{avg}}$ in the middle between the 
$\overline{x}$ values for the {ETH} and {MBL} limits. For the average gap ratio $\overline{r}$, the {slopes}  at these points are denoted by blue cross markers in Fig.~\ref{Fig:Comparison}~(a). Initially, the slope at this point is far from $\alpha_{\mathrm{max}}$ (red circles), but as the system size is increased, the slopes approach each other. Both choices exhibit the exponential scaling of $\alpha$ with the system size, as illustrated in Fig.~\ref{Fig:Comparison}~(b), featuring a similar exponent $\lambda \approx 0.1$.
Similar conclusions are drawn for the rescaled entanglement entropy $\overline{s}$ and the Schmidt gap $\Delta$ (data not shown).

{This consistency of the behavior of $\lambda$ for different choices of the point at which the slope is evaluated confirms the robustness of the exponential scaling of $|\alpha|$ at the ETH-MBL crossover in QP KIM, which is the main result of this work.} However, before we further discuss this behavior in a broader context, we analyze the system size drifts at the ETH-MBL crossover in QP KIM.

\begin{figure}[t!]
    \centering
    \includegraphics[width=1.0 \columnwidth]{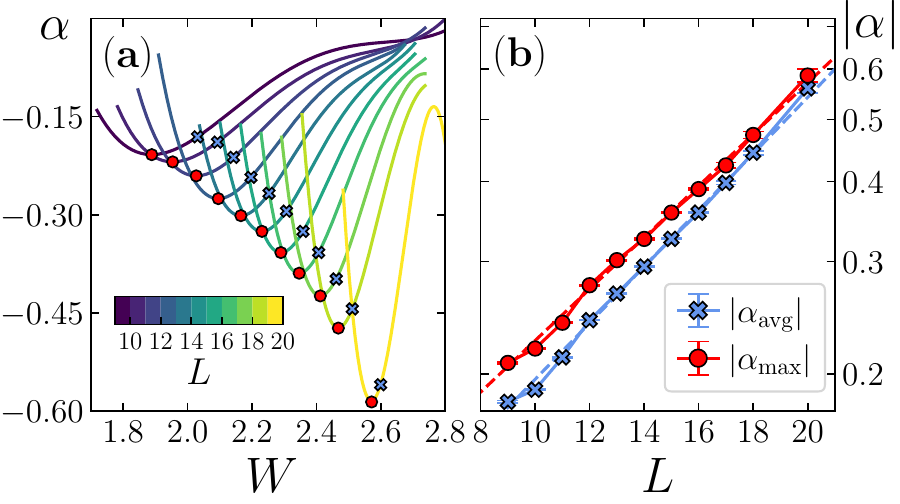}
    \caption{(a) Derivatives of $\overline{r}(W)$ for different system sizes across the MBL crossover, where the red circles represent the biggest slope and the blue cross is the slope value at $\overline{r}_{\mathrm{avg}} \approx 0.458$; (b) A direct comparison between the exponential behavior of these two different points. Despite minor differences, both curves scale with similar exponent ($\lambda \approx 0.1$).}
    \label{Fig:Comparison}
\end{figure}

\subsection{System size drifts at the {ETH-MBL} crossover}

To investigate the finite-size drifts at the ETH-MBL crossover {and shed light on possible behavior of the crossover in the large $L$ limit}, we follow \cite{Sierant20p} and consider two different size-dependent disorder strengths: 
\begin{itemize}
    \item  $W_{\overline{x}}^{T}(L)$: the disorder strength at which the ergodicity breaking indicator $\overline{x}(W)$, at a given $L$, deviates from its ergodic value by a small number $\delta_{\bar{x}}$;
    \item $W_{\overline{x}}^{*}(L)$: the crossing point between the curves $\bar{x}(W)$ for two different system sizes, $L-\Delta L$ and $L+\Delta L$, where $\Delta L \ll L$.
\end{itemize}

The former, $W_{\overline{x}}^{T}(L)$, delineates the extent of the ergodic regime at a given system size $L$. The latter, $W_{\overline{x}}^{*}$, estimates, at a given $L$, the critical disorder strength for the transition to the MBL phase. The ETH-MBL crossover observed at finite $L$ would give rise to MBL transition if $W_{\overline{x}}^{*} \stackrel{ L \to \infty }{\longrightarrow} W_c$. Conversely, if $W_{\overline{x}}^{*} \stackrel{ L \to \infty }{\longrightarrow} \infty$, there would be no stable MBL phase in the system.

{To put the results for QP KIM in a context, we recall the behavior of $W_{\overline{x}}^{*}(L)$ and $W_{\overline{x}}^{T}(L)$ in RD HSC and RD KIM.}
For the RD HSC, the disorder strength $W_{\overline{x}}^{T}$ drifts linearly with the system size,  $W_{\overline{x}}^{T} \propto L$, while the crossing point is well fitted by $W_{\bar{x}}^{*} = W_{\infty} - \mathrm{const}/L$, {for both $\overline{x} = \overline{r}$ and $\overline{x}=\overline{s}$}~\cite{Sierant20p}. One of these dependencies necessarily breaks down at $L\geq L_0^{\mathrm{HSC}} \approx 50$, the length scale at which the extrapolations of $W_{\overline{x}}^{T} $ and  $W_{\bar{x}}^{*}$ intersect. If the linear drift of $W_{\overline{r}}^{T}$ prevails, the ergodic region will grow indefinitely with system size (and consequently $W_{\bar{x}}^{*} \stackrel{L \to \infty}{\longrightarrow} \infty$), precluding the MBL phase in the thermodynamic limit. In contrast, the breakdown of linear growth $W_{\overline{x}}^{T} \propto L$ could be a premise of a stable MBL phase. Such a breakdown is not observed at currently accessible system sizes for RD HSC. However, investigations of RD KIM~\cite{Sierant23st} found that the linear drift of $W_{\bar{x}}^{T}$, observed at smaller $L$, is replaced by a slower system size dependence at $L \gtrapprox 15$. The latter observation is akin to the features of the localization crossover for the Anderson model on random graphs~\cite{Sierant23RRG}. It may suggest the presence of a stable MBL phase in RD KIM.

\begin{figure}[t!]
    \centering
    \includegraphics[width=1 \columnwidth]{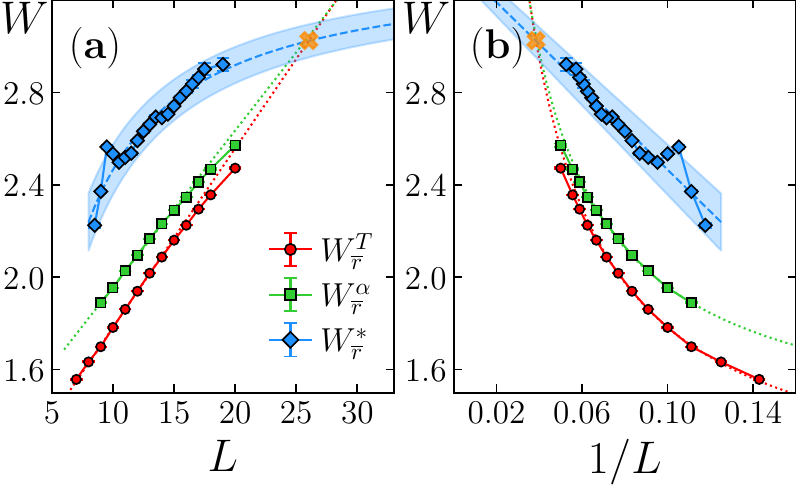}
    \caption{
    System size drifts at the ETH-MBL crossover in QP KIM. The disorder strengths $W_{\overline{r}}^{T}$, $W_{\overline{r}}^{\alpha}$, and $W_{\overline{r}}^{*}$ are plotted as functions of (a) $L$ and (b) $1/L$; the linear grow of $W_{\overline{r}}^{\alpha}$ observed for small $L$ slows down at $L\approx 16$, while $W_{\overline{r}}^{*}$, up to erratic fluctuations associated with the interplay of $L$ and $\kappa$, is well fitted by a first order polynomial in $1/L$.}
    
    \label{Fig:Crossings_QP}
\end{figure}

We start the analysis of drifts at ETH-MBL crossover in QP KIM by setting $\delta_{\bar{r}} = 0.01$ and studying the scaling of $W_{\overline{r}}^T$ with the system size. Results, represented by red circles in Fig.~\ref{Fig:Crossings_QP}, indicate a sublinear growth of $W_{\overline{r}}^T$ starting around $L \approx 16$. {The deviation from the linear increase of $W_{\overline{r}}^T(L)$ for QP KIM is weaker and occurs at a slightly larger system size than for RD KIM~\cite{Sierant23st}. Nevertheless, similarly to the RD KIM the deviation from the linear growth of $W_{\overline{r}}^T(L)$ is \textit{consistent}} with the presence of a stable MBL phase {in KIM} at sufficiently large $W$ in the $L \to \infty $ limit. 

To study the system size dependence of the crossing point $W_{\overline{r}}^{*}$, we consider $\Delta L \leq 2$. While selecting the incommensurability factor $\kappa$ from the interval centered at $\kappa_{\mathrm{gr}}$ suppresses the fluctuations of $\overline{r}$ as function of $W$ for fixed $L$, the crossing point {$W_{\overline{r}}^{*}$}   (blue diamonds in Fig.~\ref{Fig:Crossings_QP}) still exhibits signs of these fluctuations.
Analogous behavior is also present for the crossing points  {$W_{\overline{s}}^{*}$}  of the rescaled entanglement entropy curves $\overline{s}$ (data not shown), indicating that these trends are an inherent feature of models with QP potentials.
Nevertheless, $W_{\overline{r}}^{*}$ can be well approximated by a first-order polynomial in  $1/L$, as represented by the blue dashed lines in Fig.~\ref{Fig:Crossings_QP}. Extrapolating the fit $W_{\overline{r}}^{*}(L)$ to $L\to\infty$ yields $W_{\infty}\approx3.35 \pm 0.05$ which is a candidate for the critical disorder strength for MBL transition in the QP KIM.
If the \textit{linear} growth of  $W_{\overline{r}}^{T}$ and the $W_{\overline{r}}^{*}(L) \propto 1/L$ fit are extrapolated, they intersect at system size $L^{\mathrm{KIM} }_0 \approx 26$, the value similar to that obtained for RD KIM and much smaller than the one for the RD HSC. Additional technical details on finite size drifts are presented in Appendix~\ref{robustness}. {Concluding, the finite size drifts at the ETH-MBL crossover in QP KIM are qualitatively similar to the behavior observed for RD KIM~\cite{Sierant23st}. In the following, we show that the conclusion about similarities of KIM with QP and RD potentials does \textit{not} extend to the sharpness of the ETH-MBL crossover.}

\begin{figure}[t!]
    \centering
    \includegraphics[width=1 \columnwidth]{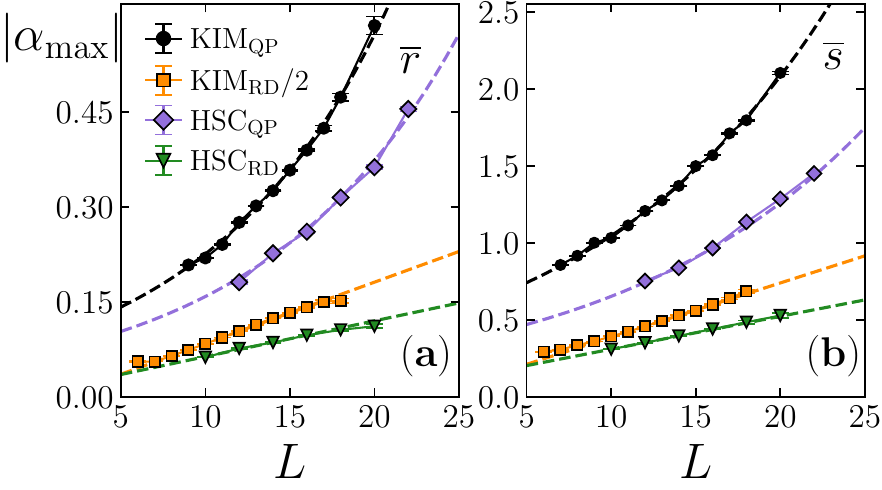}
    \caption{Sharpness of the ETH-MBL crossover in disorder spin chains. (a) The maximal slope $|\alpha_{\max}|$ of $\overline r(W)$ is shown as a function of $L$ for QP and RD KIM as well as for QP and RD HSC; dashed lines show the linear growth $|\alpha_{\max}| \propto L$ for RD models and $|\alpha_{\max}| \propto e^{\lambda L}$ for QP models (with $\lambda=0.093(4)$ for KIM and $\lambda=0.086(3)$ for HSC). The same for the rescaled entanglement entropy $\overline{s}$ is shown in (b), with $\lambda=0.068(2)$ for KIM and $\lambda=0.066(2)$ for HSC. The results for RD KIM are divided by a factor $2$ for clarity of the plot.}
    \label{Fig:Slope}
\end{figure}

\section{Sharpness of ETH-MBL crossover}

With quantitative insights into the system size drifts at the ETH-MBL crossover in QP KIM, we come back to the analysis of the sharpness of the crossover. {We consider QP and RD KIM, as well as both QP and RD HSC.} The behavior of the maximal slope $|\alpha_{\max}|$ is summarized in Fig.~\ref{Fig:Slope} both for gap ratio, panel (a), and entanglement entropy, panel (b). The data for QP KIM, denoted by black lines, highlight the substantial increase of the slopes with $L$. The system size dependence of $|\alpha_{\max}|$ is well-fitted with the exponential in $L$, and we observe no deviations from this behavior. 

The data for RD KIM, shown with orange lines in Fig.~\ref{Fig:Slope}, also reflect the growth of the slope $|\alpha_{\max}|$ with system size $L$. However, the growth is much slower and, to a good approximation, linear in $L$. This demonstrates a quantitative difference in the properties of ETH-MBL crossover in QP and RD KIM. The same conclusion is reached for the rescaled entanglement entropy $\overline{s}$, see Fig.~\ref{Fig:Slope}~(b). {This suggests a hypothesis that the growth of the maximal slope $|\alpha_{\max}|$ at the ETH-MBL crossover quantitatively differs between QP and RD systems.} 

To verify the formulated hypothesis}, we turn to HSC. 
For RD we target the $N_{ev}=\min(\mathcal{D}/100,1000)$ eigenvectors and eigenvalues closest to the mid-values of the energy spectrum $E_{m}=\frac{(E_{min}+E_{max})}{2}$ of the zero magnetization sector $\sum_j^L\sigma_j^z=0$. Here, $\mathcal{D}$ denotes the Hilbert space dimension of the HSC. For system sizes up to $L=16$, we consider at least $10^4$ disorder realizations, while for $L=18,20$, we take at least $4\cdot 10^3$ disorder realizations. The data for the QP HSC are taken from \cite{Aramthottil21}. 

\begin{figure}[t!]
    \centering
    \includegraphics[width=1.0 \columnwidth]{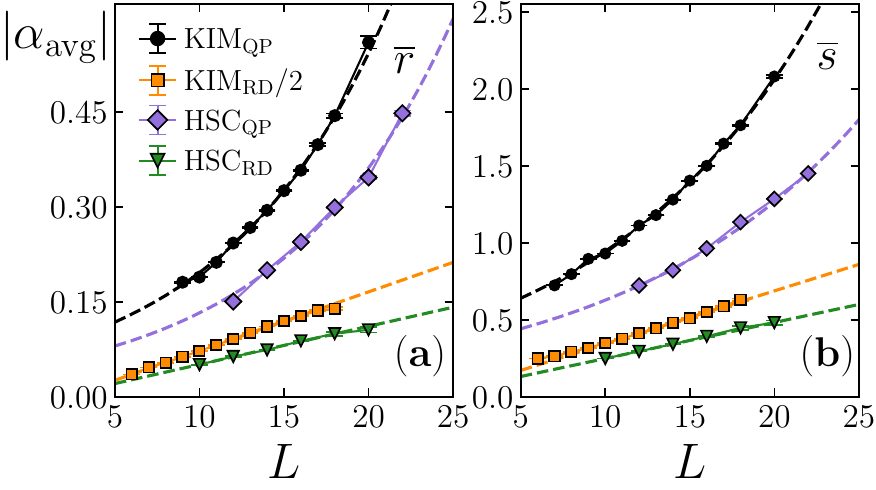}
    \caption{Similar to Fig.~\ref{Fig:Slope}, but now considering $|\alpha_{\mathrm{avg}}|$. Similarly to the maximum slope, the slopes taken at the average values $|\alpha_{\mathrm{avg}}|$ grows exponentially $|\alpha_{\mathrm{avg}}| \propto e^{\lambda L}$  for the QP models and linearly for the RD models. The exponents $\lambda$ are also similar, with $\lambda = 0.101(3)$ for the KIM and $\lambda = 0.100(4)$ for the HSC, showing the robustness of the exponential scaling.}
    \label{Fig:SlopeMid}
\end{figure}

The maximal slope $|\alpha_{\max}|$  at the ETH-MBL crossover in HSC, both for RD and QP potential, is shown in Fig.~\ref{Fig:Slope}. The quantitative difference between RD and QP results is also apparent for HSC, both for the average gap ratio data and for entanglement entropy. Indeed, for RD HSC, the maximal slope $|\alpha_{\max}|$ is increasing only linearly with system size $L$, while for QP HSC, an exponential growth of $|\alpha_{\max}|$ is found. 
Analogous conclusions, for both types of potential and both KIM and HSC, hold for the slope $|\alpha_\mathrm{avg}|$ calculated at disorder strength $W$ at which either $\overline{r}$ or $\overline{s}$ 
is equal to the average of its ergodic and localized limits, 
as shown in Fig.~\ref{Fig:SlopeMid}.

\section{Discussion}

Sharpening up of a crossover between two distinct regimes with an increase in system size is a feature consistent with an occurrence of a phase transition{in the system}.
To put our results in perspective, let us examine the consequences of a single-parameter scaling hypothesis for a transition at $W=W_c$ with a critical exponent $\nu$. According to such a hypothesis, an ergodicity breaking indicator, $\overline{x}$, can be expressed as $ {\overline{x}} = f[(W-W_c)L^{1/\nu}]$, where $f$ is a certain scaling function. Calculating the derivative and putting $W=W_c$, we find $d\overline{x}/dW = L^{1/\nu} f'(0)$, showing that the single-parameter scaling hypothesis would lead to a power-law increase of the slope $|\alpha| \propto L^{1/\nu}$, which diverges
in the $L\to\infty$ limit. The Harris criterion for RD systems~\cite{Harris74, Chayes86, Chandran15a} stipulates $\nu > 2$, predicting a slower than linear power-law increase $L^{1/\nu}$ of the slope. The data at the ETH-MBL crossover, at least at the presently accessible system sizes, \textit{is not} described by the single-parameter scaling ansatz due to the presence of the drift of the crossing point $W^*_{ \overline{x} }$~\cite{Sierant24rev}. The minimal extension of the single parameter scaling ansatz required to account for the results at ETH-MBL crossover requires the introduction of corrections to scaling in the form of sub-leading exponents~\cite{Aharony83, Slevin99} or explicit system size dependence of $W_c$~\cite{Suntajs20}. How can the findings of this work be interpreted in this context?

The slopes at the ETH-MBL crossover consistently increase for QP and RD models. We observe no deviations from this behavior. Hence, it is natural to conjecture that the slope $|\alpha_{\max}| $ keeps increasing indefinitely, leading to a non-analytic{, step function-like} behavior of the ergodicity-breaking indicators, {$\overline{x}$} in the thermodynamic limit {$L \to \infty$}. The stark difference between the exponential increase with $L$ of the slope for the QP potential and linear with $L$ growth of $|\alpha_{\max}|$ for RD case signifies that the non-analytic behavior is approached much more rapidly in QP models. However, the disorder strength $W_{\overline{r}}^{\alpha}$, at which the slope of $\overline{r}$ is maximal, follows the qualitative behavior of $W_{\overline{r}}^{T}$, and increases sublinearly with system size $L$, see Fig.~\ref{Fig:Crossings_QP}. 
{This behavior can be consistent with the presence of an MBL transition in the system if $ W_{\overline{r}}^{\alpha} \stackrel{L \to \infty}{\longrightarrow} W_c$, where $W_c$ is the critical disorder strength. In that case, the ETH-MBL phase transition occurs at $W = W_c$, the slope $|\alpha|$ diverges while the ergodicity-breaking indicators $\overline{x}$ show a abrupt jump between their values characteristic for ETH ($W<W_c$) and MBL  ($W>W_c$) phases. However, the drifts of $W_{\overline{r}}^{T}$ can also be}  consistent with the absence of MBL phase, if $W_{\overline{r}}^{\alpha} \stackrel{L \to \infty}{\longrightarrow} \infty$. Our numerical results do not exclude any of the two scenarios. This is further {supported} by a cost-function analysis of ETH-MBL crossover in QP KIM-- see Appendix~\ref{costfun}, which yields sublinear growth of the critical disorder strength, similar to QP HSC~\cite{Aramthottil21}, but milder than for RD HSC~\cite{Suntajs20}. {Nevertheless, even if $W_{\overline{r}}^{\alpha} \stackrel{L \to \infty}{\longrightarrow} \infty$, the increase of the sharpness of the ETH-MBL crossover may lead to a sharp, non-analytic change in system's properties at a disorder strength which diverges in the thermodynamic limit.}

\section{Conclusions}
In this work, we have performed a large-scale numerical investigation of ETH-MBL crossover in RD and QP spin chains. Despite the qualitative similarities in the crossover in RD and QP systems, we have demonstrated that the sharpening of the crossover with an increase in the system size is quantitatively faster in QP chains than in RD models. Our observation signals that the ergodicity-breaking phenomena in QP potential and RD systems may be essentially different, caused by different microscopic mechanisms, and may even lead to distinct system properties in the thermodynamic limit. 

The exponential sharpening of the ETH-MBL crossover in QP systems indicates that changes in the dynamics with increased QP amplitude potential are much more abrupt than in RD systems. Identification of the implications of our findings for the time evolution of local observables in QP systems is an open question. The pattern of slow oscillations of autocorrelation functions found in~\cite{Sierant22c} could be one manifestation of the sharpening up of the crossover in QP systems, observable already at intermediately long evolution times. Identifying the microscopic mechanism responsible for the differences between ETH-MBL crossover in QP and RD systems remains a high-end question that could shed new light on the status of the MBL phase.


\section{Acknowledgments}

P.R.N.F. is grateful to Konrad Pawlik for useful discussions on numerics. The work of P.R.N.F. was funded by the National Science Centre, Poland, project 2021/03/Y/ST2/00186 within the QuantERA II Programme that has received funding from the European Union Horizon 2020 research and innovation programme under Grant agreement No 101017733.  We gratefully acknowledge Polish high-performance computing infrastructure PLGrid (HPC Centers: ACK Cyfronet AGH) for providing computer facilities and support within computational grant no. PLG/2024/017289.
The work of A.S.A. has been realized within the Opus grant
 2019/35/B/ST2/00034, financed by National Science Centre (Poland). 
P.S. acknowledges support from: ERC AdG NOQIA; MICIN/AEI (PGC2018-0910.13039/501100011033, CEX2019-000910-S/10.13039/501100011033, Plan National FIDEUA PID2019-106901GB-I00, FPI; MICIIN with funding from European Union NextGenerationEU (PRTR-C17.I1): QUANTERA MAQS PCI2019-111828-2); MCIN/AEI/ 10.13039/501100011033 and by the “European Union NextGeneration EU/PRTR"  QUANTERA DYNAMITE PCI2022-132919 within the QuantERA II Programme that has received funding from the European Union’s Horizon 2020 research and innovation programme under Grant Agreement No 101017733Proyectos de I+D+I “Retos Colaboración” QUSPIN RTC2019-007196-7); Fundació Cellex; Fundació Mir-Puig; Generalitat de Catalunya (European Social Fund FEDER and CERCA program, AGAUR Grant No. 2021 SGR 01452, QuantumCAT \ U16-011424, co-funded by ERDF Operational Program of Catalonia 2014-2020); Barcelona Supercomputing Center MareNostrum (FI-2024-1-0043); EU (PASQuanS2.1, 101113690); EU Horizon 2020 FET-OPEN OPTOlogic (Grant No 899794); EU Horizon Europe Program (Grant Agreement 101080086 — NeQST),  ICFO Internal “QuantumGaudi” project; European Union’s Horizon 2020 research and innovation program under the Marie-Skłodowska-Curie grant agreement No 101029393 (STREDCH) and No 847648  (“La Caixa” Junior Leaders fellowships ID100010434: LCF/BQ/PI19/11690013, LCF/BQ/PI20/11760031,  LCF/BQ/PR20/11770012, LCF/BQ/PR21/11840013). E.P. is supported by ``Ayuda (PRE2021-098926) financiada por MCIN/AEI/ 10.13039/501100011033 y por el FSE+".
The work of J.Z.  was funded by the National Science Centre, Poland, under the OPUS call within the WEAVE program 2021/43/I/ST3/01142.
 A support by the Strategic Programme Excellence Initiative (DIGIWorkd) at Jagiellonian University is acknowledged.

Views and opinions expressed in this work are, however, those of the authors only and do not necessarily reflect those of the European Union, European Climate, Infrastructure and Environment Executive Agency (CINEA), nor any other granting authority. Neither the European Union nor any granting authority can be held responsible for them. No part of this work was written with the help of AI.

\appendix

\section{Small randomization on the incommensurability factor $\kappa$}
\label{random}

\begin{figure}[t!]
    \centering
    \includegraphics[width=0.49 \textwidth]{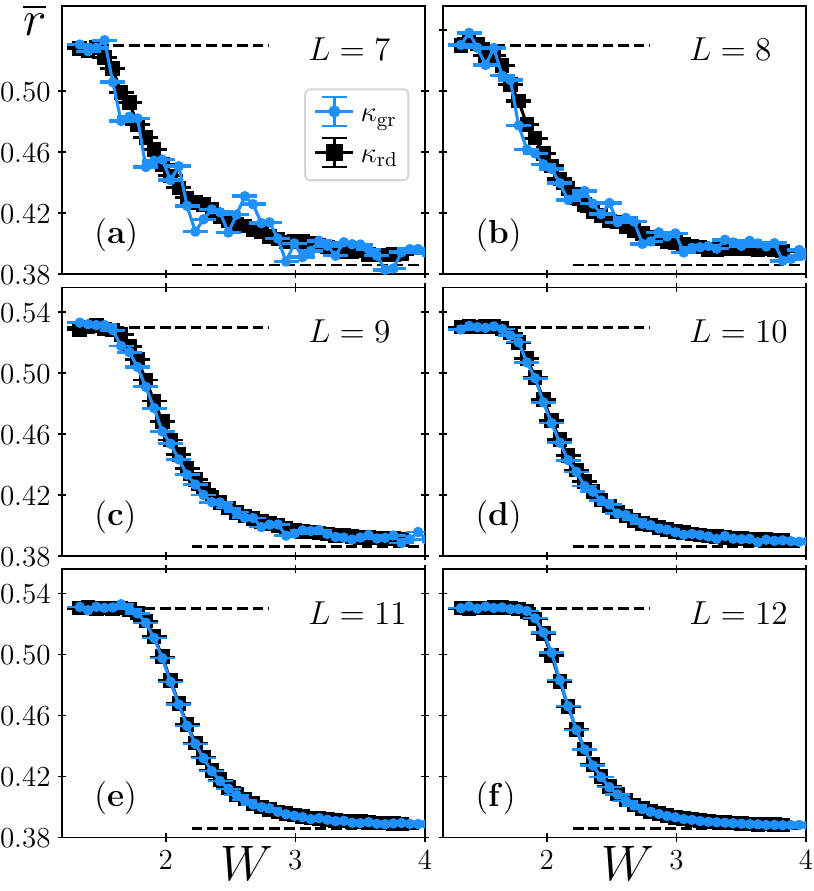}
    \caption{Average gap ratio $\overline{r}$ as a function of $W$ for different choices of $\kappa$ (see text) and system sizes $L$. Keeping the incommensurability factor $\kappa$ constant results in fluctuations, while the $\overline{r}$ curves get smoother when we randomly select $\kappa$ for each disorder realization. Both curves overlap for $L>10$.}
    \label{Fig:Gap_Small}
\end{figure}

It was mentioned in the main text that the average gap ratio $\overline{r}$ does not have a smooth behavior with $W$ for small system sizes ($L \lesssim 10$), developing fluctuations around a smooth monotonic function in $W$. Despite such length scales seeming uninteresting from the point of view of an MBL {\it phase}, they were essential in unveiling the non-trivial scaling of the finite-size drifts in the critical disorder strength $W_{\overline{x}}^{*}$ for the RD KIM \cite{Sierant23st}. 

To reduce these fluctuations and make a proper analysis on finite-size drifts of the Floquet model with QP potential, we choose $\kappa_{\mathrm{rd}} \in (0.98,1.02]\times \kappa_\mathrm{gr}$, introducing, therefore, randomness on the incommensurability factor $\kappa$. In Fig. \ref{Fig:Gap_Small}, we show a side-by-side comparison between the average gap ratio obtained with the additional randomness compared to a fixed $\kappa_\mathrm{gr}$ value. For the smallest system size ($L=7$), fluctuations between neighboring values of $W$ are pronounced when $\kappa_\mathrm{gr}$ (blue circles) is chosen, but the curve becomes smoother when we add randomness to the incommensurability factor (black squares). However, as the system size increases, the $\overline{r}$ values converge, and the curves eventually overlap for $L > 10$. Let us also mention that slight randomization of $\kappa$ does not affect in a visible way entanglement entropy statistics.

\section{Finite-size quantities}
\label{robustness}

\subsection{Finite-size drifts}

\begin{figure}[t!]
    \centering
    \includegraphics[width=1 \columnwidth]{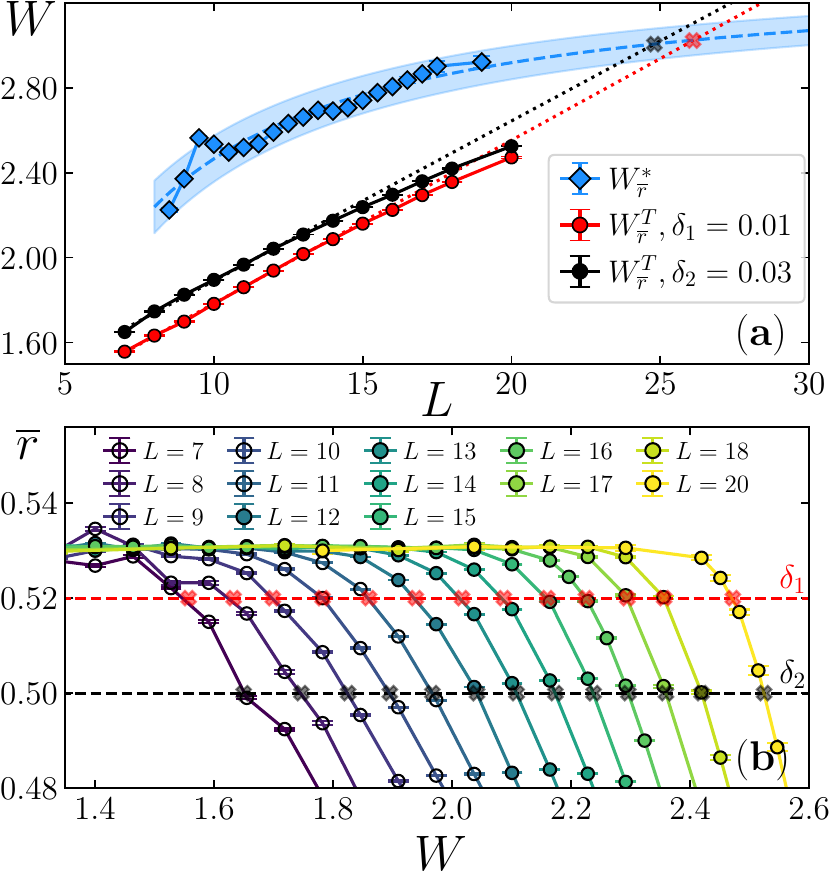}

    \caption{(a) Different choices of $\delta_{\overline{r}}$ for the near ergodic disorder strength $W_{\overline{r}}^{T}$, together with the critical point $W_{\overline{r}}^{*}$. The dotted lines represent linear scaling in $L$, while the dashed line represents a first-order polynomial in $1/L$; (b) Average gap ratio $\overline{r}$ as a function of $W$ near its COE value. The red (black) cross represents the choice of the near ergodic region for $\delta_1$ ($\delta_2$), while the filled (open) circles represent the data obtained for $\kappa_0$ ($\kappa_1$) at different system sizes. }
    \label{Dev_QP}
\end{figure}

In Fig. \ref{Dev_QP}(a), we show the details on the determination of the near ergodic disorder strength $W_{\overline{r}}^{T}$ for different choices of $\delta_{\overline{r}}$, together with the critical disorder strength $W_{\overline{r}}^{*}$ (blue diamonds). Similarly to the KIM with random disorder, the behavior of $W_{\overline{r}}^{T}$ starts to deviate from the linear scaling around $L=16$ for two different choices of $\delta_{\overline{r}}$ and this may indicate the stability of the MBL phase. The points where both values were extracted are illustrated in Fig. \ref{Dev_QP}(b), and it shows that the choice of $\delta_2 = 0.03$ leads to a faster slowdown than the $\delta_1 =0.01$ choice, consistent with the sharpness of the MBL crossover.

The critical disorder strength $W_{\overline{r}}^{*}$, despite some fluctuations that are inherent from the QP potential, can be reasonably well described by a first-order polynomial in $1/L$, with the fitting errors being shown by the shaded blue area. If we perform an extrapolation of this functional form, the crossing with the {\it linear} fitting of $W_{\overline{r}}^{T}$ happens at $L \approx 26$ ($L \approx 25$) for the choice of $\delta_1$ ($\delta_2$), as shown by the cross symbols in Fig. \ref{Dev_QP}(a).

\subsection{Slopes}

For a given $L$, we assume the observable $\overline{x}$  (which can be either $\overline{r}$ or $\overline{s}$) can be described by a polynomial function of $W$ across the entire MBL crossover. To test this hypothesis, we consider the interval $\overline{x} \in [\overline{x}_{\mathrm{COE}} - \delta_{\bar{x}},\overline{x}_{\mathrm{PS}} + \delta_{\bar{x}}]$ and fit these points with a degree $5$ polynomial function of $W$.  In our analysis, we consider $\delta_{\overline{r}} = 0.01$ for the average gap ratio and $\delta_{\overline{s}} = 0.1$ for the rescaled entanglement entropy. Except for $L=17$, all the $\chi^2$ values per degree of freedom are approximately $2.0$, suggesting that a fifth-order polynomial function in $W$ describes reasonably well the behavior of $\overline{x}$ across the transition.

We then obtain the slopes ($\alpha$) across the entire MBL crossover and extract their minimum values together with the corresponding disorder strength, $W_{\overline{x}}^{\alpha}$. We highlight the minimum value of $\alpha$ as the red circles in Fig. \ref{Fig:Comparison}(a), which clearly shows the exponential behavior of the slopes. To test the robustness of this exponential flow towards a non-analytic behavior of QP models, we also extract the slopes at the average value between $\mathrm{COE}$ and $\mathrm{Poisson}$ expectation values, $\overline{x}_{\mathrm{avg}}$. This value corresponds to $\overline{r}_{\mathrm{avg}} \approx 0.458$ if we consider average gap ratio as the quantity of interest and $\overline{s}_{\mathrm{avg}} \approx 0.5$ if we consider the rescaled entanglement entropy. In Fig. \ref{Fig:Comparison}(a), we show how both points approach each other as we increase the system size $L$, and this suggests the KIM with QP potential is very close to a step-like function behavior for $\overline{r}$. The exponential behavior of both points is shown in Fig. \ref{Fig:Comparison}(b), demonstrating the consistency of the exponential behavior, with both curves having a similar exponent ($\lambda \approx 0.093 (4)$ for the maximum and $\lambda \approx 0.101(3)$ for the average point).

To reinforce there are no qualitative differences between different choices for the slopes, we show in Fig. \ref{Fig:SlopeMid} how $\alpha$ scales if we take them at a fixed point ($\overline{x}_{\mathrm{avg}}$). The exponential behavior persists for both models with QP potential, with a similar scaling exponent of the maximum slope. For the average gap ratio, panel (a), we obtain an exponent $\lambda = 0.101(3)$ for the KIM and $\lambda = 0.100(4)$ for the HSC, while the rescaled entanglement entropy, panel (b), have exponents $\lambda = 0.078(2)$ for the KIM and $\lambda = 0.070(2)$ for the HSC. Similarly, models with random disorder exhibit a linear behavior with the system size, showing comparable scaling parameters to those observed for $|\alpha_{\mathrm{max}}|$.

\subsection{Uncertainties on finite-size quantities}
\label{Errors}

To obtain the statistical uncertainties of the observables, $\bar{r},\bar{s}$, we employ the following procedure: For a given disorder realization, we compute the average value of $x$ over $N_{\mathrm{ev}}$ different eigenstates, which we denote as $x_S$. A single disorder average itself is not self-averaged \cite{TorresHerrera20, Schiulaz20}, which may not lead to a decrease in variance $\langle (x_S - \langle x_S \rangle )^{1/2} \rangle$ as the system size increases. Assuming that $x_S$ of different disorder realizations are uncorrelated, the associated errors are calculated  as a standard error of mean over disorder realizations 
\begin{equation}
\sigma_x = \frac{(\langle (x_S - \overline{x})^2 \rangle)^{1/2}}{N_{\mathrm{dis}}^{1/2}}
\label{Std_Dev}
\end{equation}
where, $N_{\mathrm{dis}}$ is the number of disorder realizations and $\overline{x}$ denotes the average of $x_S$ over different disorder realizations.

The errors associated with each quantity $W^T_{\bar{x}},W^{\alpha}_{\bar{x}}$,$W^{*}_{\bar{x}}$ and $\vert \alpha \vert $ are then calculated using random sampling. That is, each observable, $\bar{r}$ and $\bar{s}$, is assumed to be Gaussian distributed with mean $\bar{x}$ and standard deviation $\sigma_{x}$, the quantities of interest are then repeatedly calculated over $10000$ samples from the Gaussian distributions of the observables. The mean of the distribution estimates the quantity, while the standard deviation estimates the associated errors. 

The fits made on the quantities make use of the polynomial fitting method implemented in numpy \cite{Numpy} or the curve fitting method in scipy \cite{Scipy} for the exponential fit over the slopes.

\section{Finite-size scaling analysis for QP KIM}
\label{costfun}

In the case of a phase transition, the correlation length, given as $\xi$, diverges near the critical disorder strength and acts as the dominant length scale. As a result, the normalized statistical observable, $\bar{x}$, takes a functional form 
\begin{equation}
    \bar{x}=f(L/\xi),
\end{equation}
where $f(.)$ is a continuous function. 

The two prominent proposals for how the correlation length $\xi$ diverges in the MBL transition with a random disorder is either as a standard second-order phase transition where $\xi$ diverges as a power-law fashion $\xi_0=\frac{1}{\vert W-W_c\vert^\nu }$, $\nu$ is the critical exponent and $W_c$ is the critical disorder strength based on the early renormalization group approaches \cite{Vosk15,Potter15}, or assuming a Berezinskii-Kosterlitz-Thouless (BKT) like divergence: $\xi_{B}=\exp\left\lbrace \frac{b_\pm}{\sqrt{\vert W-W_c \vert }}\right\rbrace$, where $b_{\pm}$ are arbitrary parameters on the two sides of the transition, based on the more recent real-space renormalization group approaches \cite{Goremykina19,Dumitrescu19,Morningstar19,Morningstar20} with the leading mechanism as the avalanche scenario of delocalization in the MBL phase. 

Predictions on the functional form of the correlation length are generally lacking for QP models. A few real-space renormalization group studies predict a power-law-like scaling \cite{Zhang18, Zhang2019strong}, while the absence of a mechanism akin to an avalanche scenario presently does not lead to any BKT-like predictions. However, the observation that slopes grow exponentially with increasing system size for a fixed value of $\bar{x}$ motivates us to consider both functional forms of the correlation function.

We follow the cost function minimization method introduced by Suntajs et al. \cite{Suntajs20,Suntajs20e} for the RD HSC model to best approximate the correlation length parameters. This method accurately captures the critical disorder strength for the zero-dimensional quantum sun model \cite{Suntajs22,Pawlik23}  and the finite-size drifts of critical disorder strength for the 3-D Anderson model \cite{Suntajs21}. 

Supposing the observables of interest are continuous monotonic functions of $sgn[W-W_c]L/\xi$. The cost function for a quantity $\lbrace \bar{x}_j\rbrace$ that consists of $N$ different $W$ and $L$ is defined as 
\begin{equation}
    \mathcal{C}_{\bar{x}}=\frac{\sum_{j=1}^{N-1}\vert \bar{x}_{j+1}-\bar{x}_j\vert }{\max\lbrace \bar{x}_j \rbrace-\min\lbrace \bar{x}_j \rbrace}
\end{equation}
where $\bar{x}_j$'s are sorted with respect to increasing values of $sgn[W-W_c]L/\xi$. For an ideal collapse with $\bar{x}$ being a monotonic function of $sgn[W-W_c]L/\xi$, we must have
$\sum_j \vert \bar{x}_{j+1}-\bar{x}_j\vert=\max\lbrace \bar{x}_j\rbrace-\min\lbrace \bar{x}_j\rbrace$,
and thus $\mathcal{C}_{\bar{x}}=0$. Thus, the best collapse corresponds to the global minima of $\mathcal{C}_{\bar{x}}$ for different correlation lengths and functional forms of $W_c(L)$. 

\subsection{Fixed crossover disorder strength.}
For finite-size scaling using different functional forms of correlation length, we identify the crossover disorder strength $W_c$ (we relegate the terminology from \textit{critical} to \textit{crossover} disorder strength as the existence of MBL in the thermodynamic limit is unclear) and the parameters associated with functional forms for the phase transition. We will consider observables in the interval $\bar{x}\in [\bar{x}_{COE}-\delta_{\bar{x}},\bar{x}_{PS}+\delta_{\bar{x}}]$ with $\delta_{\bar{x}}=0.03$ to avoid fluctuations once the quantity reaches the corresponding ergodic or nonergodic value. In addition, we consider system sizes $L\in [10,18]$.

The cost function is best minimized with power-law correlation function, $\xi_0$, as shown in Table \ref{table:CF_fix_wc} with values $(W_c,\nu)$ for $\bar{r} \rightarrow (2.92,0.84)$ and for $\bar{s} \rightarrow (2.82,0.876)$. While, for a BKT-like correlation function, $\xi_B$, the best $\mathcal{C}_{\bar{x}}$ minimized values $(W_c,b_+=b_-)$ are $\bar{r} \rightarrow (3.11,2.23)$ and for $\bar{s} \rightarrow (3.11,2.53)$.

\begin{table}[htb]
\centering
\caption{$\mathcal{C}_{\bar{x}}$ comparison for finite-size scaling with a fixed crossover disorder strength.}
\begin{ruledtabular}
\begin{tabular}{c c c c} 
& $\mathcal{C}_{\bar{r}}[\xi_0]$ & $\mathcal{C}_{\bar{r}}[\xi_{B}(b_+=b_-)]$ & $\mathcal{C}_{\bar{r}}[\xi_{B}(b_+\neq b_-)]$ \\ [0.5ex] 
 \colrule
 $\mathcal{C}_{\bar{r}}$& $\textbf{4.190}$& $4.515$& $4.423$\\

 \colrule
 $\mathcal{C}_{\bar{s}}$ & $\textbf{2.710}$& $3.743$& $3.439$\\
  [0.5ex] 
\end{tabular}
\end{ruledtabular}
\label{table:CF_fix_wc}
\end{table}
\subsection{Size dependent crossover disorder strength.}
To explore the possibility of system size dependent drift for the crossover disorder strength, we assume different functional dependencies on the crossover disorder strength.
As shown in Table  \ref{table:CF_drift} we find the best minimized value of cost-function is with a BKT-like functional form for the correlation length and a sub-linear drift of critical disorder strength with system size.

\begin{table}[htb]
\centering
\caption{$\mathcal{C}_{\bar{x}}$ comparison for finite-size scaling with size dependent crossover disorder strength.}
\begin{ruledtabular}
\begin{tabular}{c c c c} 
& $\mathcal{C}_{\bar{r}}[\xi_0]$ & $\mathcal{C}_{\bar{r}}[\xi_{B}(b_+=b_-)]$ & $\mathcal{C}_{\bar{r}}[\xi_{B}(b_+\neq b_-)]$ \\ [0.5ex] 
 \colrule
 $W_0+W_1L$ & $0.954$& $1.252$& $1.080$\\
 $W_0+W_1\ln (L)$ & $0.976$& $\textbf{0.465}$& $\textbf{0.465}$\\
 $W_0+W_1/{L}$& $2.076$& $1.852$& $1.852$\\
 $W_0+{W_1}/{[\ln(L)]}$ & $0.975$& $0.925$& $0.925$\\
 $W_0+W_1L^{\gamma}$ & $0.934$& $0.480$& $0.479$\\[0.5ex] 
\end{tabular}
\end{ruledtabular}
\begin{ruledtabular}
\begin{tabular}{c c c c} 
& $\mathcal{C}_{\bar{s}}[\xi_0]$ & $\mathcal{C}_{\bar{s}}[\xi_{B}(b_+=b_-)]$ & $\mathcal{C}_{\bar{s}}[\xi_{B}(b_+\neq b_-)]$ \\ [0.5ex] 
 \colrule
 $W_0+W_1L$ & $1.166$& $1.871$& $1.833$\\
 $W_0+W_1\ln (L)$ & $1.169$& $1.063$& $1.063$\\
 $W_0+W_1/{L}$ & $2.324$& $2.343$& $2.343$\\
 $W_0+{W_1}/{[\ln(L)]}$& $1.170$& $1.278$& $1.278$\\
 $W_0+W_1L^{\gamma}$ & $1.160$& $\textbf{1.054}$& $\textbf{1.042}$\\[0.5ex] 
\end{tabular}
\end{ruledtabular}
\label{table:CF_drift}
\end{table}

With a correlation function $\xi_B(b_-=b_+)$ and the functional form for drifts in critical disorder strength as $W_0+W_1 \ln(L)$ we find the best-minimized parameters $(W_0,W_1,b)$ for $\bar{r} \rightarrow(-0.91,1.09,1.23)$ while for $\bar{s} \rightarrow (-1.48,1.16,0.48)$. For a functional form for drifts in critical disorder strength $W_0+W_1L^{\gamma}$ the best minimized parameters $(W_0,W_1,\gamma,b)$ for $\bar{r} \rightarrow (-6.24,5.95,0.12,1.19)$ and for $\bar{s} \rightarrow (-5.68,5.03,0.14,2.32)$. Considering $b_-\neq b_+$, we find no difference in values except in the case of $b_-$, which shows a substantial change. 

\begin{figure}[t!]
    \centering
    \includegraphics[width=1.0 \columnwidth]{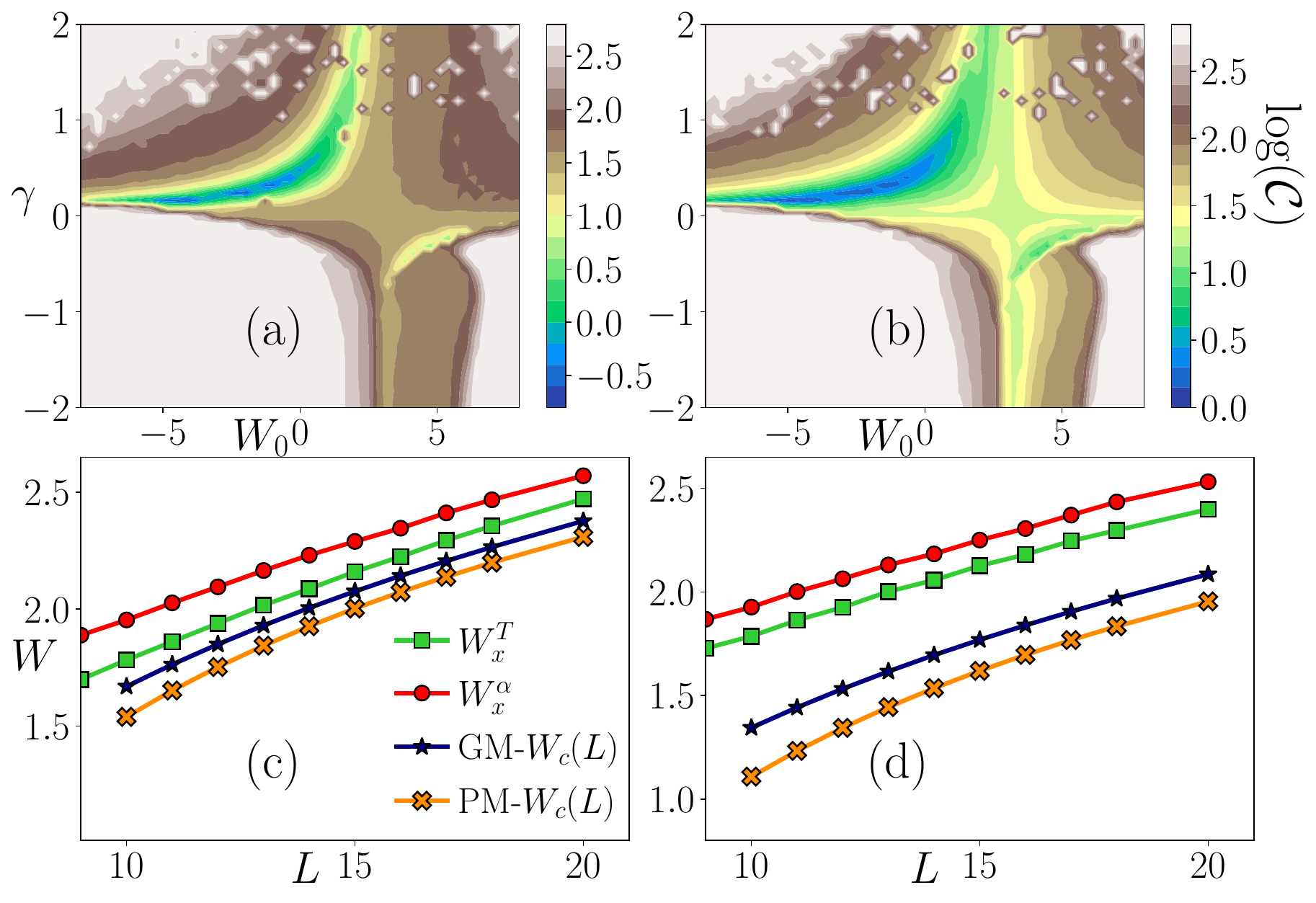}
    \caption{Assuming a BKT-like correlation function $\xi_B$ and a functional form for drifts $W_c(L)=W_0+W_1L^\gamma$ (a) and (b) shows how the logarithm of the cost-function changes in the ($\gamma$,$W_0$) space for $\bar{r}$ and $\bar{s}$ respectively. While (c) and (d) make a comparison of the minimized $W_c(L)$, both for a global (GM) and physical (PM) minima, with disorder strengths $W_{\bar{x}}^T$ and $W_{\bar{x}}^{\alpha}$ for $\bar{r}$ and $\bar{s}$ respectively.}
    \label{Fig:BKT_with_pd}
\end{figure}

In the parameter space of $\mathcal{C}_{\bar{x}}$ with a BKT-like correlation function with system size drifts in crossover disorder strengths $W_c(L)=W_0+W_1L^\gamma$ over $(W_0,\gamma)$ as shown in Fig. \ref{Fig:BKT_with_pd} we find an island with global minima (GM) that extends to $W_0 < 0 $. The parameter values resulting from these minima can be non-physical, leading to a negative critical disorder strength for small values. We also find an island with physically relevant minima (PM) that has $W_0 > 0$ and $-1<\gamma<0$. The parameters $(W_0,W_1,\gamma,b)$ found with PM are for $\bar{r}\rightarrow (7.34,-9.31,-0.21,1.29)$
while for $\bar{s}\rightarrow (6.94,-9.82,-0.22,2.40)$. Indicating the system size-dependent drift of the crossover disorder strength saturating at $W\approx 7$. 

\subsection{Numerical details on cost-function }
To minimize the cost function, we employ the differential evolution method implemented in SciPy \cite{Scipy}. We run a series of independent differential evolution algorithms with different random seeds for each $\mathcal{C}_{\bar{x}}$. We employ $1000$ independent realizations to find the optimal minimization. In each realization, we allow up to $10^4$ iterations with a relative tolerance of convergence $10^{-4}$ and a population size of $250$.

%


\end{document}